# Development of a NbN Deposition Process for Superconducting Quantum Sensors

D. M. Glowacka, D. J. Goldie, S. Withington, H. Muhammad, G. Yassin, *Member, IEEE*, and B. K. Tan

*Abstract*— We have carried out a detailed programme to explore the superconducting characteristics of reactive DC-magnetron sputtered NbN. The basic principle is to ignite a plasma using argon, and then to introduce a small additional nitrogen flow to achieve the nitridation of a Nb target. Subsequent sputtering leads to the deposition of NbN onto the host substrate. The characteristics of a sputtered film depend on a number of parameters: argon pressure, nitrogen flow rate and time-evolution profile, substrate material, etc. Crucially, the hysteresis in the target voltage as a function of the nitrogen flow can be used to provide a highly effective monitor of nitrogen consumption during the reactive process. By studying these dependencies we have been able to achieve highly reproducible film characteristics on sapphire, silicon dioxide on silicon, and silicon nitride on silicon. Intrinsic film stress was minimised by optimising the argon pressure, giving NbN films having $T_c$ = 14.65 K. In the paper, we report characteristics such as deposition rate, Residual Resistance Ratio (RRR), film resistivity, transition temperature, and stress, as a function of deposition conditions. In order to enhance our understanding of the microwave properties of the films, we fabricated a wide range of microstrip NbN resonators (half wavelength, quarter wavelength, ring resonators). In the paper, we provide an illustrative result from this work showing a 2.1097 GHz resonator having a Q of 15,962 at 3.3 K.

## I. Introduction

THE superconducting characteristics and stoichiometry of Niobium Nitride (NbN) make it a potentially important material for fabricating a wide variety of quantum sensors and microcircuits. It can have a critical temperature ($T_c$) as high as 15K, leading to Kinetic Inductance Detectors (KIDs) operating with bath temperatures of around 4 K., and a pair breaking gap frequency of 1.2 THz, making it interesting for Superconductor-Insulator-Superconductor (SIS) mixers operating at frequencies of potentially up to 2 THz.

Clearly it is essential to be able to control the properties of the films, such as critical current, surface roughness, grain size and film stress, but NbN is more difficult to deposit than Nb because of the metastable [1] superconducting phase of the material, and because the films must be deposited by reactive sputtering. A particular difficulty is a hysteretic effect that occurs with the addition of the reactive gas during sputtering, which in turn affects stoichiometry. It is very important to quantify the location of the best operating point, and this requires a careful study of the effects of deposition conditions. Thakoor et al. [2] found that high-Tc NbN films were deposited at the point where the total sputter pressure started to increase rapidly with increasing nitrogen flow. In this paper, we outline work that we have done to create a high-yield highly-reproducible deposition process for high-Tc NbN. We also briefly describe work on realising and measuring resonators of the kind that would be suitable for a KID array operating at 3.5 K in an easy-to-use closed-cycle refrigerator.

## II. Experimental Procedure

NbN films were deposited in an ultra-high vacuum, DC planar magnetron sputtering system using a 3″ Nb target (99.9% pure), with argon (99.9999% pure) and nitrogen (99.9999% pure) as the sputter gases. The flow rate of each gas was controlled individually by using precision mass-flow controllers. The NbN films were deposited at room temperature onto sapphire and also onto 200nm silicon nitride membranes on silicon substrates. Typically, the base pressure prior to sputtering was in the low $10^{-9}$ or high $10^{-10}$ Torr.

The complete hysteresis curve of the target voltage had to be determined as the nitrogen flow rate was first increased and then decreased; this curve enables the best operating point to be found [3]. First we pre-sputtered the target for 4 minutes at the desired argon partial pressure to remove any target surface contamination. The nitrogen flow rate was then increased at increments of 0.5 sccm to a point well above target nitridation (about 5.5 sccm in practice). After saturation of the target, the reverse branch of the hysteresis curve was determined by decreasing the nitrogen flow rate by the same increments. A series of NbN depositions were made with constant power and argon pressure at room temperature, using a nitrogen flow rate in the range 0.6 to 1.6 sccm. To ensure that the films were deposited on the proper branch of the hysteresis curve, the nitrogen flow was first raised well above the saturation point of the target for 2 minutes, and then decreased to the desired flow. The Nb target was pre-sputtered for 4 minutes in only argon gas before each NbN deposition. The NbN thickness, Residual Resistance Ratio (RRR), room temperature resistivity (RT) and 20 K resistivity ($\rho_{20}$) were measured for all samples. The $T_c$ was only measured for the samples deposited with a flow rate in the downward sloping region of the hysteresis curve.

D. M. Glowacka, D. J. Goldie, S. Withington and H. Muhammad are with the Department of Physics, University of Cambridge, JJ Thomson Avenue, Cambridge CB3 0HE, UK. E-mail: glowacka@mrao.cam.ac.uk
G. Yassin and B-K. Tan are with the Departmant of Physics, University of Oxford, Denys Wilkinson Building, Keble Road, Oxford OX1 3RH, UK

The fabrication of multilayer devices not only requires NbN with high $T_c$, reproducible $T_c$, and low resistivity, but also low film stress. To investigate stress, we deposited NbN on 50 mm diameter discs of 0.008 mm thick polyimide (Kapton) and measured the resulting radii of curvature of the samples. First, we deposited NbN with the conditions needed to give the highest $T_c$, then we measured the stress, and then we deposited NbN with the same nitrogen flow, but adjusted the Ar pressure to achieve films with minimal compressive stress.

Once the optimum deposition conditions had been determined, we fabricated thin-film microwave resonators, of the kind that could be used for KIDs, to establish whether the RF conductivity of the films was high. Prior to every resonator processing run, we deposited calibration samples having thicknesses of around 100 nm. These were used to check that high-quality films were indeed being laid down.

The microwave resonators were processed on 76.20±0.03 mm diameter and 300±25 μm thick silicon wafers, with 1μm thermal oxide layers on both sides. Both sides of the wafer were polished. The processing started with depositing a 200 nm NbN ground plane, which was patterned using a Reactive Ion Etcher (RIE). 550 nm of silicon dioxide was then deposited as the insulator between the ground plane and the 200 nm thick NbN resonator tracks. The last step involved the deposition of 200nm silicon dioxide to form the coupling capacitors, and finally a 200 nm layer of NbN for the microstrip wiring.

### III. RESULTS

The target voltage hysteresis curve for an argon flow rate of 16.2 sccm, giving a partial pressure of 3.4 mTorr, and constant target power of 100 W is shown in Fig. 1.

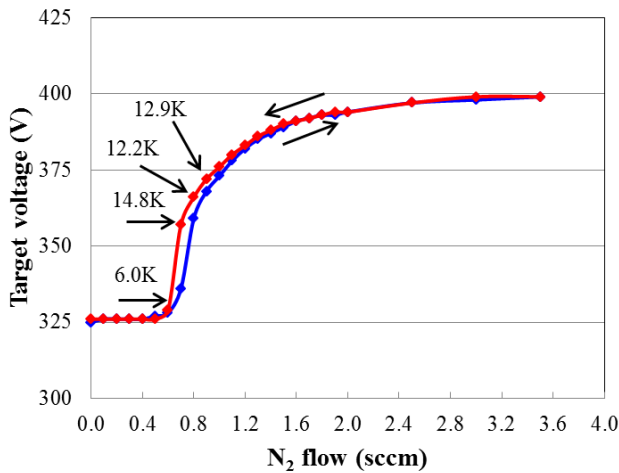

Fig. 1. The target voltage as a function of nitrogen flow with a constant argon flow of 16.2 sccm and power of 100 W. The $T_c$'s of NbN films deposited at the four operating points on the hysteresis curve are shown.

The nitrogen consumption was monitored by measuring the total process pressure. In Fig. 2 the total pressure is plotted against the nitrogen flow rate. Hedbabny and Rogalla [4] pointed out that NbN films with optimum critical temperature occur at nitrogen flow rates close to the transition point, which in our case was at 3.5 mTorr and 0.7 sccm.

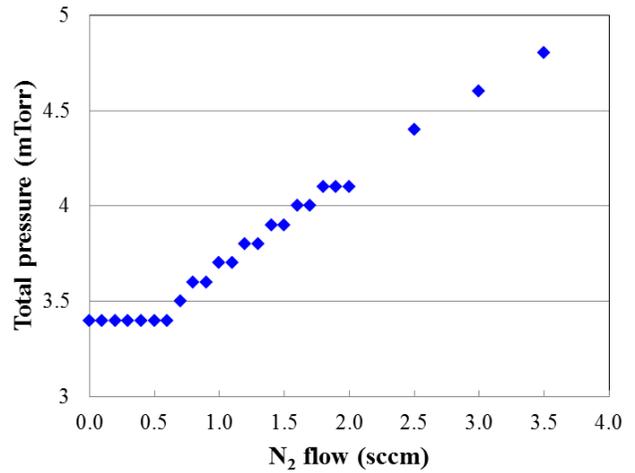

Fig.2. Total process pressure versus the nitrogen flow rate in the reactive sputtering process.

Fig. 3 shows the niobium target voltage as a function of flow rate on the decreasing branch of the hysteresis for argon pressures of 3.4 mT and 5.0 mT; again, the sputtering power was 100 W.

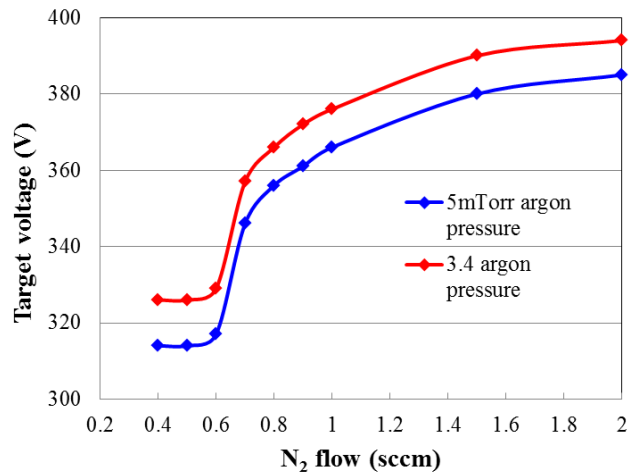

Fig.3. Niobium target voltage as a function of flow rate, on the decreasing branch of the hysteresis curve for argon pressures of 3.4 mT and 5.0 mT

A series of NbN films were deposited on sapphire and on silicon nitride films on silicon wafers at operating points located on the decreasing slope of hysteresis curve with an argon partial pressure of 3.4 mT. The nitrogen flow was varied from 0.6 to 1.6 sccm, and we measured the resulting film thickness, RT resistivity, $\rho_{20}$ resistivity and RRR. The results are shown in Fig. 4, Fig. 5 and Fig. 6 respectively.

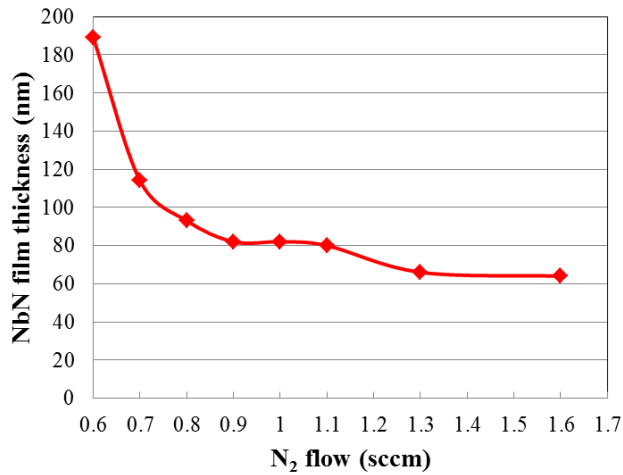

Fig.4. NbN film thickness as a function of nitrogen flow rate.

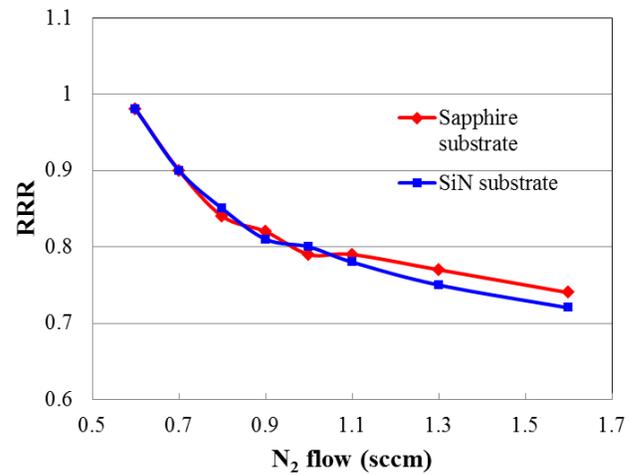

Fig.6. Residual Resistivity Ratio (RRR) as a function of nitrogen flow rate.

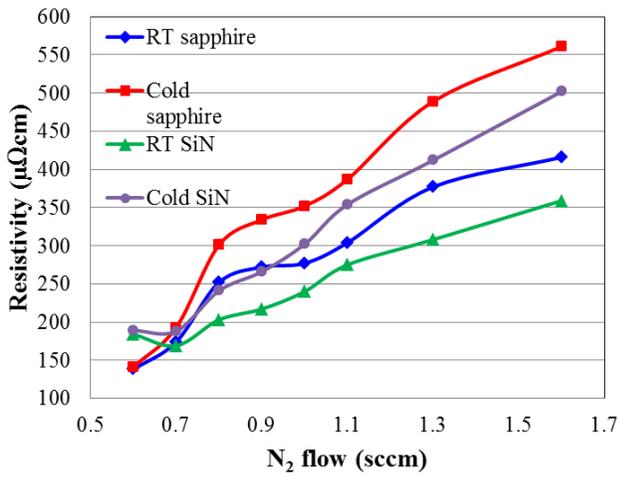

Fig.5. Room temperature (RT) and 20 K (cold) resistivity as a function of nitrogen flow rate.

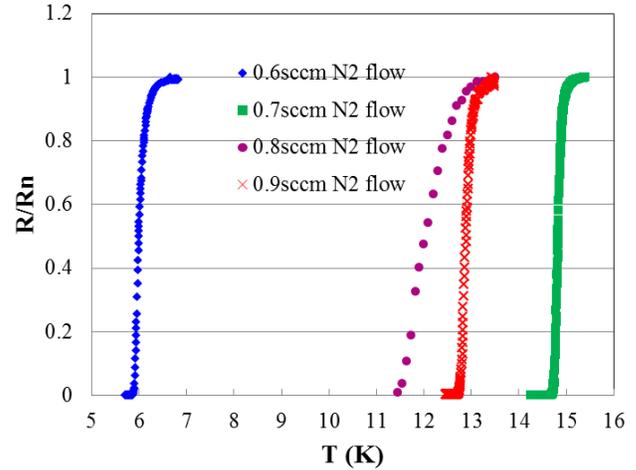

Fig.7. Critical temperatures of NbN films deposited at 3.4 mTorr argon pressure on silicon nitride substrates.

The resistance-temperature curves resulting from 4 different nitrogen flow rates, 0.6, 0.7, 0.8 and 0.9 sccm, are shown in Fig. 7. The highest $T_c$ defined as the temperature where R/Rn=0.5, 14.83K, was measured for a film deposited using a nitrogen flow rate of 0.7 sccm. We deposited two more NbN films on sapphire substrates and silicon nitride substrates with a 0.7 sccm nitrogen flow. The variation of resistivity on the sapphire substrates was ±3% and on the silicon-nitride substrates was ±4.5%. The RRR values were very similar for all depositions at around 0.9 and 0.89 respectively.

The Stoney equation was used to calculate film stress from knowledge of the film thickness, substrate thickness, and radius of curvature of the deformed substrate. The calculated stress for a NbN film deposited using a 0.7 sccm nitrogen flow, 100W power and total pressure of 3.5 mTorr was compressive at about 1100-1600 MPa. The argon pressure is the biggest contributor to total sputtering pressure, and so to minimize the compressive stress we estimated that the argon pressure should be around 5 mTorr. The stress for a NbN film deposited with a 0.7 sccm nitrogen flow rate at 5 mTorr was a smaller tensile stress of about 650MPa. The argon pressure was then reduced to 4.5 mTorr (0.7 sccm nitrogen flow) to further reduce the tensile stress. The resulting film had very small tensile stress about 130 MPa. The stress is shown as a function of argon partial pressure in Fig. 8.

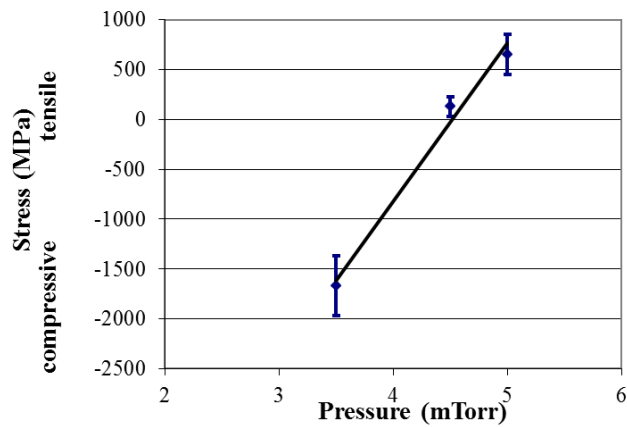

Fig.8. Film stress for argon deposition pressures of 3.4 mT, 4.5 mT and 5mT.

Measured resistance-temperature plots for films deposited on silicon nitride with 0.7 sccm nitrogen flow and argon pressures of 3.4 (two separate deposition runs), 4.3 and 4.5 mTorr are shown in Fig.9.

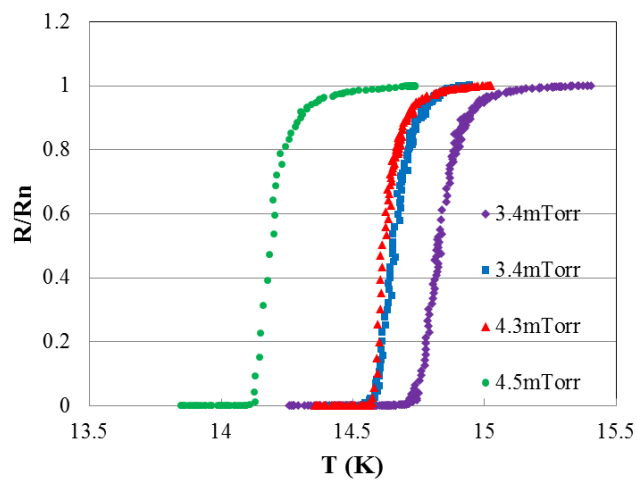

Fig.9. Resistance-temperature plots for films deposited with different argon pressures. The measured Tc's were 14.83 K and 14.65 K at 3.4 mTorr, 14.63 K at 4.3 mTorr and 14.19 K at 4.5 mTorr argon partial pressure.

The film deposited with a 4.3 mT argon partial pressure had a high critical temperature, 14.63 K, and very low compressive stress.

A series of NbN films were then deposited keeping the argon partial pressure constant at 4.3 mTorr, but using nitrogen flow rates of 0.7, 0.68 and 0.65 sccm. We measured films thickness, resistivity, RRR and $T_c$, and the results of the resistance-temperature measurements are shown in Fig. 10.

Immediately prior to fabricating microwave resonators, we deposited calibration samples to check the repeatability of the characteristics. The films were deposited at 4.3 mTorr argon partial pressure, 0.7 sccm nitrogen flow and 100W target power. The results are shown in Fig. 11.

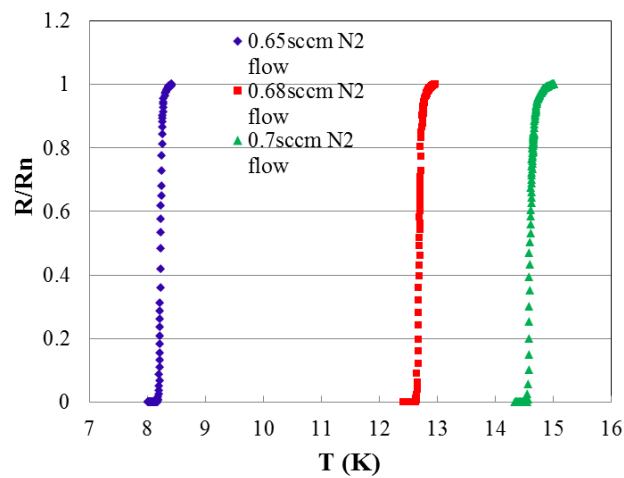

Fig.10. Resistance versus temperature for NbN films deposited with different nitrogen flow rates. The Tc for 0.65 sccm was 8.25 K, for 0.68 sccm was 12.69 K and for 0.7 was 14.63 K. The films with the highest Tc had thickness 124 nm, RT resistivity 229 µΩcm and an RRR of 0.82.

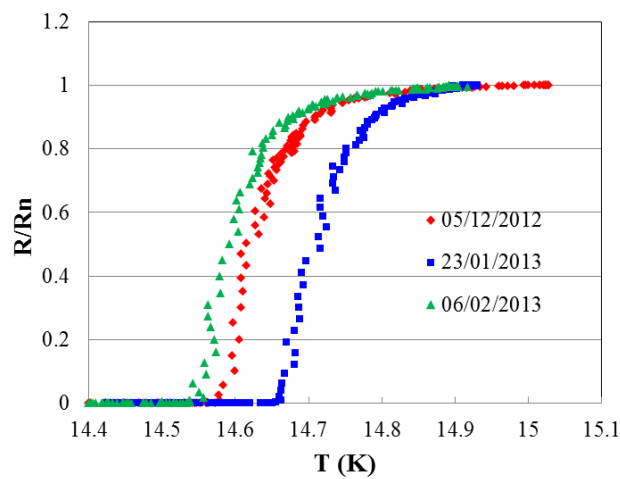

Fig. 11. Normal resistance verses temperature for films deposited with notionally identical conditions. The measured $T_c$'s were 14.59K (green), 14.63K (red) and 14.67K (blue). The RT resistivity and RRR were 217 µΩcm and 0.84 (green), 229 µΩcm and 0.82 (red), 237 µΩcm and 0.81 (blue).

Much thinner films (<50nm) were also investigated. Using the same deposition conditions as Fig. 11, 6 nm, 12 nm, 24 nm and 48 nm films were deposited and characterised. The measured $T_c$'s and $R_{sq}$'s are shown in Fig. 12. Because the normal resistance increases with decreasing film thickness, we assume that the grain size also decreases, and that the change in $T_c$ merely reflects the changing grain size.

A key part of our work was to investigate whether the deposited films have low-loss microwave properties, which would make them suitable for KID imaging arrays at 3.5 K. We constructed detailed electromagnetic models, based on the complex conductivity of BCS superconductors and the equations of Yassin and Withington [5], of half-wavelength, quarter-wavelength and ring resonators. Coupling to the readout lines was achieved by thin-film parallel-plate capacitors fabricated using sputtered silicon dioxide as the dielectric.

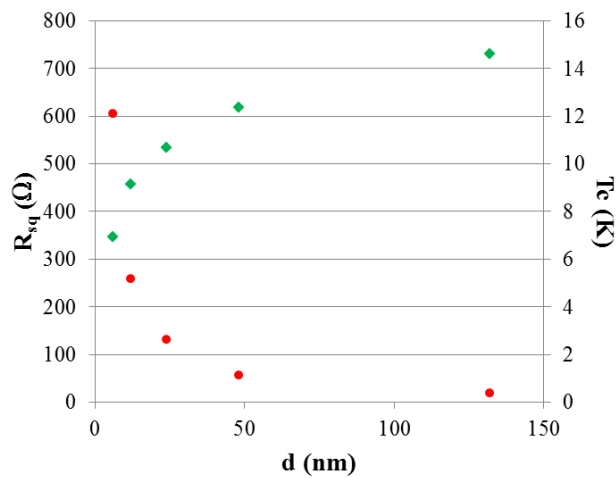

Fig. 12. Square resistance and transition temperature as a function of film thickness: 6 nm, 12 nm, 24 nm and 48 nm films were used.

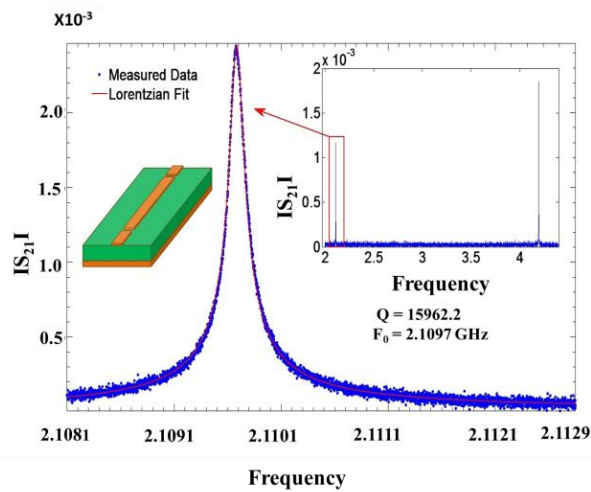

Fig. 13. Transmission response of λ/2 NbN microstrip resonator at 3.3K.

The transmission parameters of the devices were measured at 3.3 K in a closed-cycle refrigerator, and fitted to Lorentzian line profiles. The resonator mask also had test structures to allow us to measure the critical temperatures of the NbN films deposited during different stages of device processing. The $T_c$'s of two different NbN depositions were 14.64 K and 14.76 K. In Fig. 13 we show a resonance curve (blue) and Lorentzian fit (red) of one of our half-wavelength microstrip resonators. The insert, over a wider frequency range, also shows a resonance when the structure is one wavelength long. The first resonant frequency was measured to be 2.1097 GHz with a Q-factor of 15,962, the second resonance was at 4.194 GHz with a Q factor of 10,469, and the third resonance was at 6.285 GHz with a Q factor of 8,665, all at 3.3 K. Our electromagnetic simulation predicted a resonance frequency of 2.4389 GHz and a Q factor of 25,700.0. Given the uncertainties in the electromagnetic modelling of the precise geometry, this is a pleasing result, and demonstrates that high-Q resonators can be fabricated for use at 3.5 K. We have also devised a procedure for extracting the complex conductivity of the material from measured resonance curves. In the case of Fig. 13, we measured a complex conductivity of $\sigma = 0.5 \times 10^4 + i\, 2.27 \times 10^8\, (\Omega\, m)^{-1}$, which compares with the BCS calculated result $\sigma = 1.14 \times 10^4 + i\, 3.34 \times 10^8\, (\Omega\, m)^{-1}$.

## IV. CONCLUSION

We have developed an optimized reactive sputtering process for high-quality high-Tc NbN films, and we have successfully used the technique to fabricate multilayer micro-resonators. Crucially, the process has been shown to work on sapphire, silicon nitride on silicon, and silicon dioxide on silicon. Care is required to ensure that the target is fully nitrided prior to film deposition commencing. The highly-reproducible, stable, low-stress high-Tc process was achieved by measuring film characteristics for various nitrogen flow rates and argon pressures. Our preferred route uses a 4.3 mT argon partial pressure, a 0.7 sccm nitrogen flow rate, and a 100W target power. The films deposited with these conditions had room temperature resistivities between 193 and 23 μΩcm, resistivities at 20 K between 220 and 293 μΩcm, RRR 0.81 and 0.88, and $T_c$'s between 14.59 and 14.67 K. With these films we were able to make half-wavelength resonators having Q's of around 16,000 at 2 GHz when operated in a closed-cycle fridge at 3.3 K.